\documentclass[%
superscriptaddress,
preprint,
 amsmath,amssymb,
 aps,
]{revtex4-1}

\usepackage{graphicx}
\usepackage{dcolumn}
\usepackage{bm}

\usepackage{color}

\begin{document}


\title{Infrared spectroscopic study of carrier scattering in gated CVD graphene}

\author{Kwangnam Yu}
\affiliation{Department of Physics, University of Seoul, Seoul 130-743, Korea}

\author{Jiho Kim}
\affiliation{Department of Physics, University of Seoul, Seoul 130-743, Korea}

\author{Joo Youn Kim}
\affiliation{Department of Physics, University of Seoul, Seoul 130-743, Korea}

\author{Wonki Lee}
\affiliation{Institute of Advanced Composites Materials, Korea Institute of Science and Technology (KIST), Wanju 55324, Korea}

\author{Jun Yeon Hwang}
\affiliation{Institute of Advanced Composites Materials, Korea Institute of Science and Technology (KIST), Wanju 55324, Korea}

\author{E. H. Hwang}
\affiliation{SKKU Advanced Institute of Nanotechnology and Department of Physics, Sungkyunkwan University, Suwon 16419, Korea}

\author{E. J. Choi}
 \email{Corresponding author, e-mail : echoi@uos.ac.kr, Tel : +82-2-6490-2649, Fax : +82-2-6490-2644}
\affiliation{Department of Physics, University of Seoul, Seoul 130-743, Korea}

\date{\today}

\begin{abstract}
We measured Drude absorption of gated CVD graphene using far-infrared transmission spectroscopy,
and determined carrier scattering rate ($\gamma$) as function of the varied carrier density ($n$).
The $n$-dependent $\gamma(n)$ was obtained for a series of conditions systematically changed as 
(10~K, vacuum) $\rightarrow$ (300~K, vacuum) $\rightarrow$ (300~K, ambient pressure), 
which reveals that (1) at low-T, charged impurity ($=A/\sqrt{n}$) and short-range defect ($=B\sqrt{n}$) 
are the major scattering sources which constitute the total scattering $\gamma=A/\sqrt{n}+B\sqrt{n}$,
(2) among various kinds of phonons populated at room-$T$,
surface polar phonon of the SiO$_{2}$ substrate is the dominantly scattering source, 
(3) in air, the gas molecules adsorbed on graphene play a dual role in carrier scattering as charged impurity center
and resonant scattering center.
We present the absolute scattering strengths of those individual scattering sources, 
which provides the complete map of scattering mechanism of CVD graphene for the first time.
This scattering map  allows us find out practical measures to suppress the individual scatterings, 
the mobility gains accompanied by them, and finally the ultimate attainable carrier mobility for CVD graphene. 
\end{abstract}

\maketitle


\section*{Introduction}

Two dimensional (2D) graphene has been extensively studied during recent years 
because of its fundamental interest and device applications. 
The latter one was highlighted with particular attention by the successful synthesis of the large scale graphene 
by use of the chemical vapor deposition (CVD) method.\cite{kim2009large,bae2010roll,lee2010wafer,li2009large}
The CVD graphene with wafer-scale size, often as large as 30-inch, \cite{bae2010roll} 
opened a route to practical application of graphene as various devices such as flexible display.
However, despite the great merit in the large size, 
CVD graphene exhibits carrier mobility which is definitely low compared with that of the exfoliated graphene,
indicating that the Dirac electron (hole) experiences more scattering in the former one.
For the CVD graphene supported by substrate, the carrier is scattered by various kinds of scattering sources. 
To name some of them,   
impurity and defect are created during the growth, transfer, and device fabrication of graphene.
Pseudomagnetic field due to random strain produces another scattering channel to carrier. \cite{Cuoto} 
The acoustic- and optical-phonons are thermally populated in the graphene and also in the substrate,  
which scatter the carrier through 
the electron-phonon interaction.\cite{hwang2008acoustic,zou2010deposition,chen2008intrinsic} 
Furthermore, when the graphene device is operated in ambient condition, air molecules
adhere to graphene and scatter the carrier.\cite{robinson2008adsorbate,wehling2009adsorbates} 

In order to improve the mobility of CVD graphene, a thorough knowledge must be gained 
on the carrier scattering mechanism. 
Specifically one should identify all the scattering sources participating in the carrier scattering, 
determine their precise scattering strengths in terms of well-defined quantity such as 
the scattering rate, find out the practical ways of removing or reducing the scattering sources, 
and finally, based on these investigations, should present the highest mobility value that can be achieved 
in the scattering-suppressed CVD graphene.  
To this end, extensive experimental efforts must be devoted, which has been lacking to date. 
To perform such experiment one should take special care to choose a measurement tool that can effectively
probe the carrier scattering and mobility.  
In particular, because the carrier mobility $\mu$ is determined by the carrier density $n$ 
and scattering rate $\gamma$ in graphene as $\mu \sim 1/(\gamma\sqrt{n})$, 
it is desirable that the tool is capable of detecting $n$ and $\gamma$ simultaneously in a single measurement.
Also, noting that various kinds of scatterings are mixed together in graphene sample, 
one should devise a strategy to disentangle them and measure the individual scattering strengths separately. 

In this work we perform a Far-infrared (FIR) transmission experiment of CVD graphene supported on 
SiO$_{2}$/Si substrate.   
The FIR spectroscopy, a remote and contactless probe of carrier dynamics, measures Drude response of the Dirac carrier.
The strong merit of measuring the Drude peak is that $\gamma$ and $n$ are measured simultaneously, 
allowing us figure out the relationship between them, $\gamma(n)$. 
We show that  the scattering function $\gamma(n)$, obtained for a range of $n$ by electrostatically gating the device,  
provides the key to identify the scattering sources and determine their scattering strengths.

\section*{Experiment}

A CVD-grown graphene, 5~mm by 5~mm in size, was transferred on SiO$_{2}$/Si substrate \cite{li2009large} 
and the sample was mounted in a temperature-variable (10~K~$<T<$~300~K) transport/optical dual purpose cryostat 
and treated by thermal annealing (350~$^{o}$C, 1.5~h) in vacuum (10$^{-6}$ Torr).
The transport $I$-$V$ curve and FIR optical transmission were measured for the same sample 
as gating voltage $V_{G}$ was applied to graphene for the hole-doping and then electron-doping regime.  
A commercial FTIR (Bomem DA8) and He-cooled bolometer were used for the FIR measurement. 
We used the RefFit-program to rigorously analyze FIR-data for quantitative analysis\cite{kuzmenko2005kramers}.

\section*{Results and discussion}

Figure 1(a) shows $T(V_{\rm{G}},\omega)$ taken at $T=10$~K and $P=10^{-6}$~Torr. 
Gating voltage was applied into the hole-doping regime $V_{\rm{G}}<0$~V.
The free carrier absorption becomes stronger as more carrier is induced in graphene with increasing $|V_{\rm{G}}|$.
We calculate the optical conductivity $\sigma_{1} (\omega)$ by applying the Kramers-Kronig constrained variational
method to $T(V_{\rm{G}}, \omega)$ \cite{kuzmenko2005kramers}.
For quantitative analysis we fit $\sigma_{1} (\omega)$ using the Drude model
\begin{equation}
\sigma(\omega)=\frac{\sigma_{\rm{DC}}}{1+i \cdot \frac{\omega}{\gamma}}
\end{equation}
where DC-conductivity $\sigma_{\rm{DC}}$ and carrier scattering rate $\gamma$ are the fitting parameters.
The fitting curves show good agreement with data [Fig. 1(b)]. 
In the inset we compare the FIR-$\sigma_{\rm{DC}}$ with transport $\sigma_{\rm{DC}}$ where
the latter is extracted from $I$-$V$ data 
using the relation $\sigma_{\rm{DC}}=(L/W)(I_{\rm{SD}}/V_{\rm{SD}})$, $L$ = channel length and $W$ = channel width.
The two results are in good agreement, confirming the consistency between our transport and optical experiments.
The $I$-$V$ curve shows that the charge neutral voltage is very low ($V_{\rm{CNP}}=2.5$~V), 
the curve is fairly symmetric for the hole- and electron-doping, and 
the carrier mobility is reasonably high ($\mu=5400$~$\rm{cm}^{2}/\rm{Vs}$ at $V_{\rm{G}}=-20$~V), 
demonstrating high quality of our sample. 
FIR-data for the electron-doping regime is presented in Supplementary, Fig. S1. 

Fig. 2(a) shows how $\gamma$ changes with $V_{\rm{G}}$. 
Here we plot $\sigma_{1}(\omega)/\sigma_{\rm{DC}}$, the $\sigma_{1}(\omega)$ normalized by $\sigma_{\rm{DC}}$.
$\gamma$ corresponds to the curve width measured at half maximum $\sigma_{1}(\omega)/\sigma_{\rm{DC}}=1/2$. 
$\gamma$ is large at low doping and decreases as $V_{\rm{G}}$ increases. 
To double-check this behavior of $\gamma$, we seek another way of data analysis: 
From Eq. (1), it follows that    
\begin{equation}
\frac{1}{\sigma_{1}(\omega)}=a\omega^{2}+b
\end{equation} 
where $a=1/(\sigma_{\rm{DC}} \cdot \gamma^{2})$, $b=1/\sigma_{\rm{DC}}$.
The linear dependence of $1/\sigma_{1}(\omega)$ on $\omega^{2}$ is confirmed in Figure 2(b). 
By performing linear-fit of data, we obtain the slope ($=a$) and $x$-intercept ($=b$), 
from which $\sigma_{\rm{DC}}$ and $\gamma$ are calculated.  
They reproduce the results of Fig.2(a). 
For later comparison of data with theory, we convert the $\gamma$.vs.$V_{\rm{G}}$ relation into $\gamma$.vs.$n$. 
For this end we calculate carrier density $n$ for each $V_{\rm{G}}$ using the relation 
$Q=CV_{\rm{G}}$ where $Q=qn$, $C=\epsilon_{\rm{r}}\epsilon_{0}A/d$.
Note that $n$ can be measured alternatively from the Drude weight and also from interband optical transition (Supplementary Fig. S3).

In Fig. 3 we show $\gamma(n)$. 
The electron-doping regime is included in this plot.
For graphene impurity and defect are inevitably introduced during the growth and they produce major scattering 
at low temperature like in other solids. 
For a charged-impurity (CI) $\gamma(n)$ decreases 
as $n$ increases due to screening of the Coulomb potential, $\gamma_{\rm{CI}}$ = $A$/$\sqrt{n}$ 
\cite{sarma2011electronic,hwang2007carrier}.
On the other hand, charge-neutral defects are often described by  
the short-ranged (SR) potential and the scattering for this case increases as   
$\gamma_{\rm{SR}}=B\sqrt{n}$ for 2D graphene \cite{sarma2011electronic,hwang2007carrier}.
We assume that they coexist  
\begin{equation}
\gamma=\gamma_{\rm{CI}}+\gamma_{\rm{SR}}=\frac{A}{\sqrt{n}}+B\sqrt{n}
\end{equation} 
and use Eq. (3) to fit our $\gamma(n)$.
The fit agrees reasonably well with data for hole-doping regime. 
To double-check the data fit we multiply $\sqrt{n}$ to Eq. (3) 
which leads to   
$\gamma\sqrt{n}=A+Bn$. 
The linear behavior of $\gamma\sqrt{n}$ with respect to $n$ is confirmed in the data (inset), 
and the coefficients $A (=5.0 \times 10^{7}$~$\rm{cm}^{-2}$, the $y$-intercept) and $B (=1.2 \times 10^{-5}$, slope) 
are determined.
For the electron-doping regime the data deviates from the fit, which indicates small amount of air or ice inevitably adsorbed on graphene at low-T (see Supplimentary VI for details).
Eq. (3) predicts that $\gamma(n)$ should {\it increase} when $n$ is very high due to the $\gamma_{\rm{SR}}$ term. 
Recently we measured the FIR-$\gamma$ up to 3 $\times$ 10$^{13}$~$\rm{cm}^{-2}$
using the electronic double layer (EDL) gating and found that indeed $\gamma$ increases as predicted.     
In the 2D scattering theory $A$ and $B$ are given as $A \approx 6\times10^{-4} n_{\rm{CI}}$ 
and $B \approx 2.8 \times 10^4  n_{\rm{SR}}V_0^2$, where $n_{\rm{CI}}$ and $n_{\rm{SR}}$ are the density of the CI 
and the SR respectively, and $V_{0}$ is the potential strength of the SR. 
Comparison with our experimental $A$ and $B$ shows that $n_{\rm{CI}} \approx 10^{11}$ cm$^{-2}$ 
and $n_{\rm{SR}} V_0^2 = 4.33$ (eV\AA)$^2$.  
As for $n_{\rm{SR}}$, the exact potential strength $V_{0}$ is not known $n_{\rm{SR}}$ can not be determined. 
However, $n_{\rm{SR}}V_0^2$ of our sample is about 10 times stronger than 
the $n_{\rm{SR}}V_0^2$ of high mobility exfoliated graphene \cite{hwang2007carrier}
indicating that SR is significant in CVD graphene.

Next we increase the sample temperature to $T=300$~K and repeat the $V_{\rm{G}}$-controlled FIR measurement. 
Figure 4 shows that $\gamma$ has increased compared with $T=10$~K. 
We calculate $T$-driven scattering rate $\Delta \gamma = \gamma(300\rm{K}) - \gamma(10\rm{K})$
which decreases with increasing $n$ (red curve). 
At room-$T$, phonons are populated in graphene and substrate.
In graphene, acoustic phonons (AP) and optical phonons (OP) create the phonon-carrier scattering 
with the rates $\gamma_{\rm{AP}} \approx 6 \sqrt{\tilde{n}}$ cm$^{-1}$ 
($\tilde{n}$ is the $n$ measured in $10^{12}$ cm$^{-2}$) and $\gamma_{\rm{OP}} \approx 
\gamma_{\rm{0}}$ (independent of $n$) respectively \cite{hwang2008acoustic}. 
Obviously they do not account for $\Delta \gamma$ in terms of the $n$-dependence. 
As for the SiO$_{2}$ substrate, optical phonon in conjunction with the polar-discontinuity
produces the surface polar phonon (SPP) scattering to graphene.
Because the SPP has the nature of the long-range Coulomb interaction, it is screened by carrier in graphene 
and  $\gamma_{\rm{SPP}}$  decresaes with increasing $n$. In Ref. 16 authors calculated 
$\gamma_{\rm{SPP}}(n)$  to plot other quantaties such as the dc-conductivity. 
We compared it  with our $\Delta\gamma$ which showed that they agree 
in their $n$-dependent behaviors, i.e, the sharp increase of scattering with $n \rightarrow 0$ in the vicinity of $n$=0, 
leading to the conclusion that  SPP of the substrate is the main scattering source among other  temperature-induced scatterings.

In ambient condition air molecules adhere to graphene and become a scattering sources.
To study this scattering effect, we purged the sample chamber with air at $T=300$~K.
The $I$-$V$ curve shifts to positive $V_{\rm{G}}$ along with the elapse of time, indicating that   
graphene is progressively doped by hole [inset of Fig. 5(a)].
After the $I$-$V$ shift has saturated 
we measured FIR-$\gamma$ and calculated the adsorbate scattering (AD) 
$\gamma_{\rm{AD}} \equiv \gamma(\rm{air})-\gamma(\rm{vacuum})$.
Figure 5(a) shows that $\gamma_{\rm{AD}}$ is asymmetric for the hole-doping and electron-doping. 
We propose that $\gamma_{\rm{AD}}$ consists of two components 
as $\gamma_{\rm{AD}} = \gamma_{\rm{CI}} + \gamma_{\rm{X}}$
where $\gamma_{\rm{CI}}$ is the CI-scattering $\gamma_{\rm{CI}} = D/\sqrt{n}$ as we discussed above, 
and $\gamma_{\rm{X}}$ is a scattering of unknown origin at this point. 
The $I$-$V$ shift (or hole-doping) implies that Dirac electron in graphene moves to the adsorbed gas. 
The adsorbate is then negatively charged, behaving as a CI-scatterer to graphene. 
For hole-doping regime $V_{\rm{G}}$ $<$ $V_{\rm{CNP}}$, 
$\gamma_{\rm{AD}}$ can be fit with the CI-scattering alone [Fig. 5(b)], i.e, 
$\gamma_{\rm{AD}}=D/\sqrt{n}$ ($D=4.2\times10^{7}$~cm$^{-2}$) and $\gamma_{\rm{X}}=0$.
For the electron-doping $V_{\rm{G}}$ $>$ $V_{\rm{CNP}}$ non-zero $\gamma_{\rm{X}}$ is needed in addition 
to $\gamma_{\rm{CI}}$ to fit data. 
$\gamma_{\rm{X}}=\gamma_{\rm{AD}}-D/\sqrt{n}$ shows the $n$-dependent shape of this scattering (black curve).
Robinson {\it et al.} studied theoretically the orbital hybridization of graphene and adsorbate, 
revealing that carrier in graphene undergoes a resonant behavior when scattered by the hybridized 
orbital \cite{robinson2008adsorbate}.
They showed that the resonant scattering $\gamma_{\rm{RS}}$ is highly asymmetric for the hole and electron carrier. 
For instance we reproduce one of their results in the inset of Fig. 5(b), 
which is almost zero for hole and strong for electron, consistent with our experimental $\gamma_{\rm{X}}$. 
Based on this observation we assign $\gamma_{\rm{X}}$ to $\gamma_{\rm{RS}}$ and propose that 
adsorbate plays the dual role as Coulomb scatterer and the resonant scatterer.

In Fig. 6 we plot all $\gamma(n)$'s determined in this work in one graph.
It shows quantitatively how the total scattering 
$\gamma = \gamma_{\rm{CI}} + \gamma_{\rm{SR}} + \gamma_{\rm{SPP}} + \gamma_{\rm{AD}}$ 
is composed in terms of the individual scatterings.
For example, at a moderate hole-doping $n=-2\times10^{12}$~cm$^{-2}$, the four scatterings contribute to 
$\gamma$ in $1: 0.48 : 0.53 : 0.84$ ratio. 
For electron-side $n=2\times10^{12}$~cm$^{-2}$ the ratio changes to $1 : 0.48 : 0.76 : 2.24 $.
In graphene the carrier mobility is determined from $n$ and $\gamma$, 
$\mu=(v_{\rm{F}}e/\sqrt{\pi}\hbar)(1/\gamma\sqrt{n})$ where the Fermi velocity is $v_{\rm{F}}=1.1\times10^{8}$~cm/s. 
For the FET operated at $T=300$~K and in air, $\gamma=100.5$~cm$^{-1}$ from Fig. 6 and we have
$\mu=3500$~cm$^{2}$/Vs (we take $n=-2\times10^{12}$~cm$^{-2}$ again). 
How can one improve the mobility to higher value? 
If we assume the impurity and defect could be reduced to the quality of exfoliated graphene 
through the advances of growth technique, 
the single-crystal CVD graphene being one recent example of them \cite{lee2014wafer}, 
then $\gamma_{\rm{CI}} + \gamma_{\rm{SR}}$ can be reduced from current value 52.4~cm$^{-1}$ to 
$\sim$ 17.7~cm$^{-1}$ \cite{tan2007measurement}. 
In terms of the total $\gamma$, 
$\gamma$ is reduced from 100.5~cm$^{-1}$ to 65.8~cm$^{-1}$ and $\mu$ increases to 5340~cm$^{2}$/Vs. 
For further improvement we note that $\gamma_{\rm{SPP}}$ can be reduced by employing a substrate 
with high energy optical phonon.
For example hexagonal boron nitride (hBN) has optical phonon at 
$\hbar\omega_{\rm{OP}}\sim170$~meV \cite{geick1966normal}, higher than $\hbar\omega_{\rm{OP}}\sim 59$~meV of SiO$_{2}$. 
As result the number of thermal populated SPP, $n_{\rm{SPP}}\sim e^{-\hbar\omega_{\rm{OP}}/k_{\rm{B}}T}$,  
decreases by 1/72 of that for SiO$_{2}$. 
If we assume $\gamma_{\rm{SPP}}$ is proportional to $n_{\rm{SPP}}$ it is reduced 
from $\gamma_{\rm{SPP}}=18.4$~cm$^{-1}$ (Fig. 6) to $\gamma_{\rm{SPP}}=0.3$~cm$^{-1}$. 
The total scattering decreases to $\gamma$ = 47.7~cm$^{-1}$ leading to the enhanced mobility $\mu=7400$~cm$^{2}$/Vs.
Finally, as for $\gamma_{\rm{AD}}$, the gas adsorption can be prevented by covering graphene 
with a proper passivation material.  
In fact hBN is an efficient passivator as proved for black phosphorus (BP) 
and MoS$_{2}$\cite{doganov2015transport,lee2015highly}. 
If we assume that few-layer hBN is integrated with graphene it will suppress 
the adsorption scattering $\gamma_{\rm{AD}}$ at the cost of another hBN "substrate" scattering 
$\gamma_{\rm{SPP}}$ introduced from the top. 
For the hBN(top)/graphene/hBN(substrate)/Si total scattering 
$\gamma = \gamma_{\rm{CI}} + \gamma_{\rm{SR}}$ + 2 $\times$ hBN-$\gamma_{\rm{SPP}}$ (2 = top + bottom) 
is reduced to 18.3~cm$^{-1}$, because hBN-$\gamma_{\rm{SPP}}<\gamma_{\rm{AD}}$ (see Fig. 6), and
the mobility is increased to $\mu=19300$~cm$^{2}$/Vs.
We regard this value is probably the highest mobility one can practically achieve for room-$T$ 
and in-air FET operation.   
To make the suggested multi-stack structure, synthesis of high quality CVD hBN and integration technique of 
two CVD material sheets are necessary, which were demonstrated recently \cite{jang2016wafer, kim2016chiral}.   
   
To conclude, we have obtained complete map of scattering mechanism of CVD graphene 
by performing FIR-transmission measurements.
The four scattering sources were identified and their individual scattering strengths were measured.
This quantitative and source-resolved scattering map (Fig. 6) provides insight into improved carrier mobility by 
(1) removing the impurity and defect in graphene and substrate,
(2) inserting hBN layer between graphene and SiO$_{2}$ substrate,
(3) protecting graphene from air using hBN passivation, 
where the ultimate mobility is estimated to be $\mu=19300$~cm$^{2}$/Vs for practical device operation 
at room-T and in air.
Our findings offer invaluable guide to the application of the large scale graphene as high-speed electronic device.   


\begin{figure*}[ht]
\centering
\includegraphics[width=0.7\linewidth]{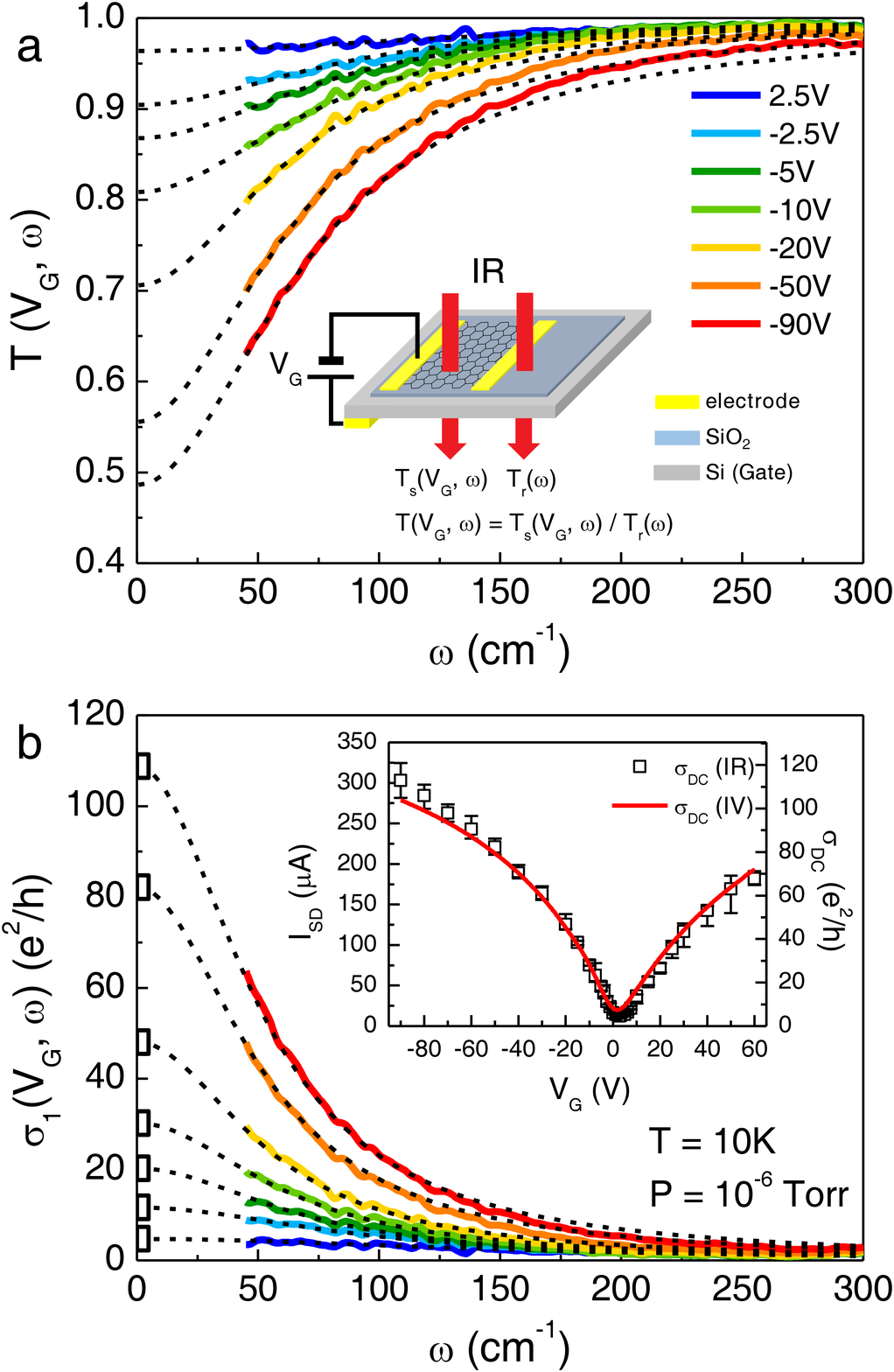}
\caption{
(a) FIR transmission $T(V_{\rm{G}}, \omega)$ measured with the gating voltage applied negatively $V_{\rm{G}}<0$~V. 
Inset shows schematically how the $V_{\rm{G}}$-dependent transmission measurement is made. 
(b) Optical conductivity $\sigma_{1}(V_{\rm{G}}, \omega)$ calculated from $T(V_{\rm{G}}, \omega)$. 
Dashed curve shows the Drude-model fitting result. 
The squared symbol at $\omega=0$~$\rm{cm}^{-1}$ shows the DC-conductivity $\sigma_{\rm{DC}}$(IR). 
In the inset we compare the $\sigma_{\rm{DC}}$ calculated from the $I$-$V$ curve with 
$\sigma_{\rm{DC}}$(IR).
}
\label{fig1}
\end{figure*}

\begin{figure*}[ht]
\centering
\includegraphics[width=0.7\linewidth]{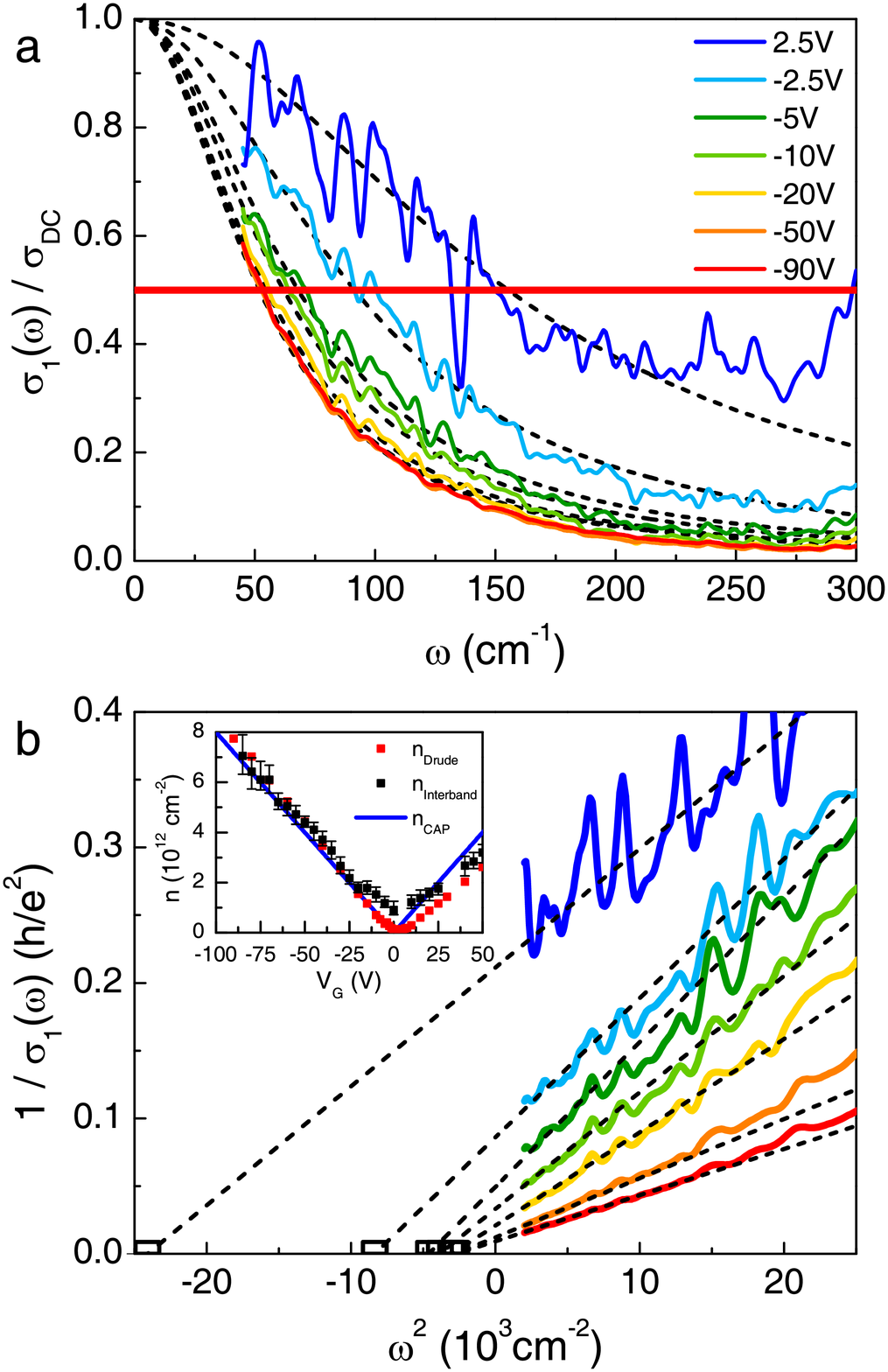}
\caption{
(a) Optical conductivity $\sigma_{1}(\omega)$ is normalized by the DC-conductivity $\sigma_{\rm{DC}}$. 
The curve width measured at $\sigma_{1}(\omega)/\sigma_{\rm{DC}}=1/2$ represents the scattering rate $\gamma$. 
(b) Plot of $1/\sigma_{1}(\omega)$ versus $\omega^{2}$.
Dashed lines are the linear fit using the Drude relation $1/\sigma_{1}(\omega) = a \omega^{2}+ b$.
The $x$-axis intercept (squared symbol) shows -$\gamma^{2}$ determined for each $V_{\rm{G}}$. 
Inset shows carrier density $n$ measured from the $y$-axis intercept and from two other methods (see Supplementary Figure S3).
}
\label{fig2}
\end{figure*}

\begin{figure*}[ht]
\centering
\includegraphics[width=0.7\linewidth]{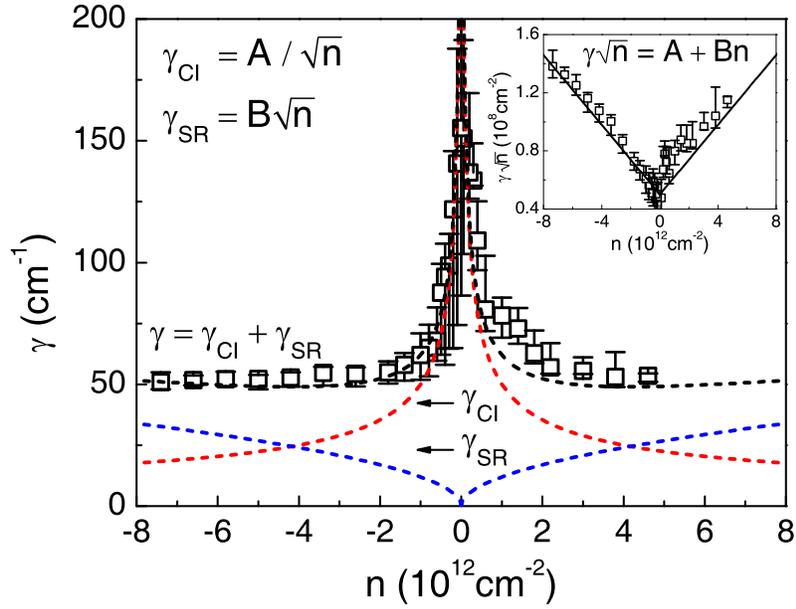}
\caption{
Scattering rate $\gamma$ plotted against carrier density $n$ for the hole-doping ($n<0$~cm$^{-2}$) 
and electron-doping ($n>0$~cm$^{-2}$).
The dashed curves show $\gamma=\gamma_{\rm{CI}}+\gamma_{\rm{SR}}$ where $\gamma_{\rm{CI}}=A/\sqrt{n}$ 
and $\gamma_{\rm{SR}}=B\sqrt{n}$ represent the Coulomb scattering and short range scattering rate, respectively.
Inset shows the linear relation $\gamma\sqrt{n} = A + Bn$. 
}
\label{fig3}
\end{figure*}

\begin{figure*}[ht]
\centering
\includegraphics[width=0.7\linewidth]{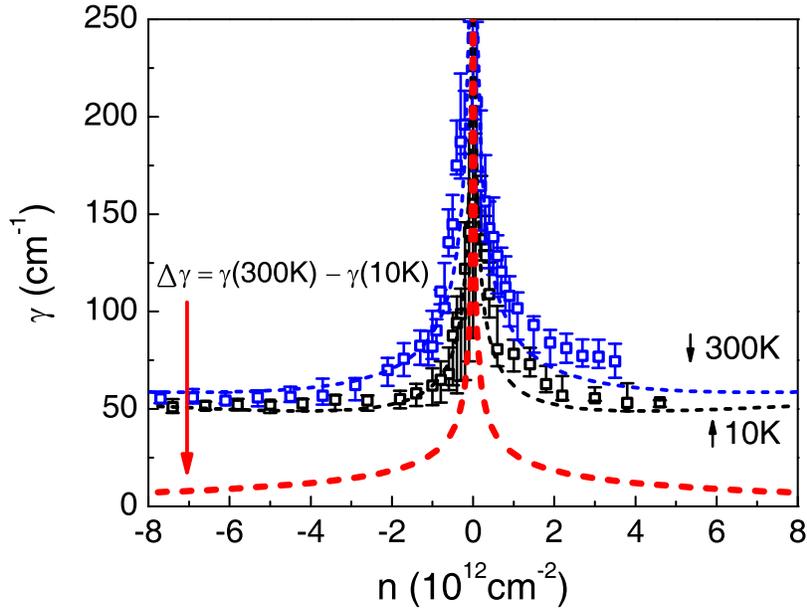}
\caption{
Scattering function $\gamma(n)$ measured at $T=300$~K and $P=10^{-6}$~Torr. 
The temperature-induced scattering $\Delta\gamma \equiv \gamma(300\rm{K})-\gamma(10\rm{K})$ is shown by the red curve.
}
\label{fig4}
\end{figure*}

\begin{figure*}[ht]
\centering
\includegraphics[width=0.7\linewidth]{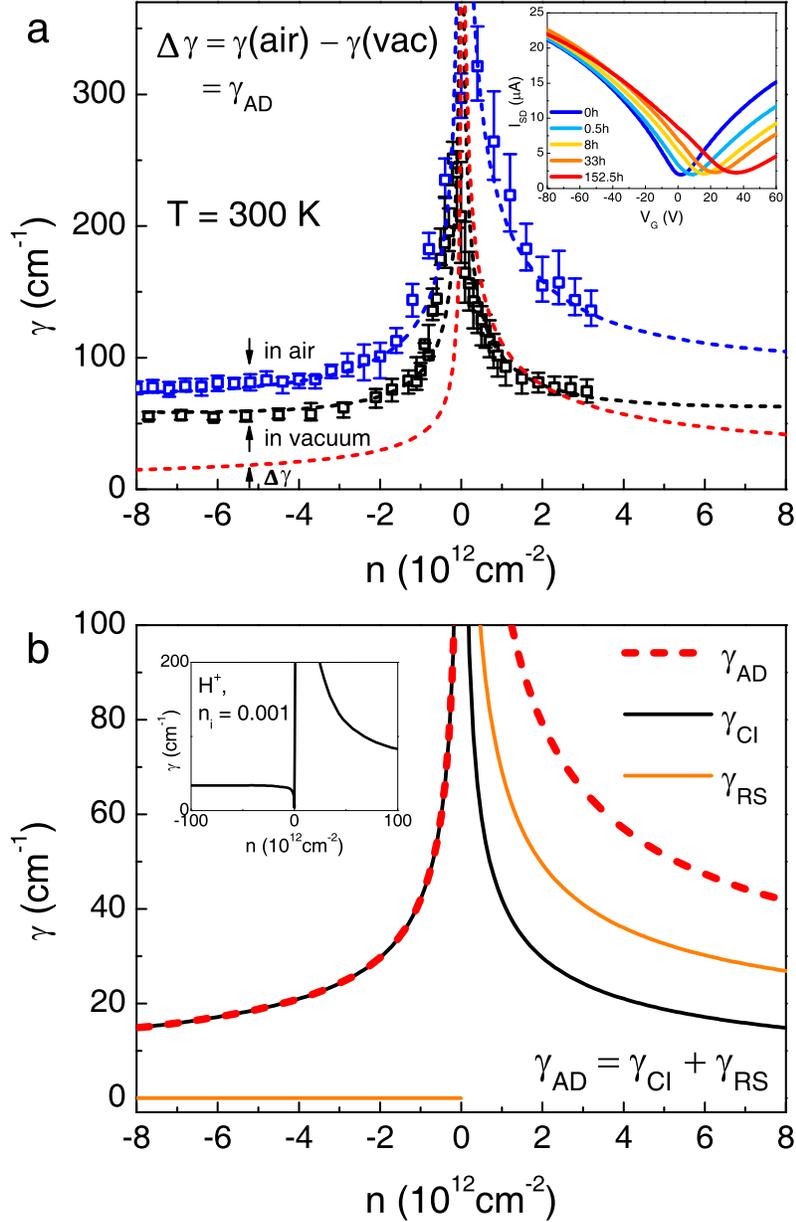}
\caption{
(a) $\gamma$ measured before and after graphene is exposed to air.
The difference $\Delta\gamma=\gamma(\rm{air})-\gamma(\rm{vacuum})$ 
shows the scattering due to the gas adsorbed on graphene, $\gamma_{\rm{AD}}$. 
Inset shows the evolution of the $I$-$V$ curve with the time elapsed since exposure. 
(b) $\gamma_{\rm{AD}}$ is analyzed in terms of two components $\gamma_{\rm{AD}}= \gamma_{\rm{CI}} + \gamma_{\rm{RS}}$ 
where $\gamma_{\rm{CI}}$ and $\gamma_{\rm{RS}}$ are the Coulomb scattering and resonant scattering, respectively.   
In the inset we reproduce the theoretical $\gamma_{\rm{RS}}$ calculated for H$^{+}$- adsorbed graphene with permission 
from the authors\cite{robinson2008adsorbate}.
}
\label{fig5}
\end{figure*}

\begin{figure*}[ht]
\centering
\includegraphics[width=0.7\linewidth]{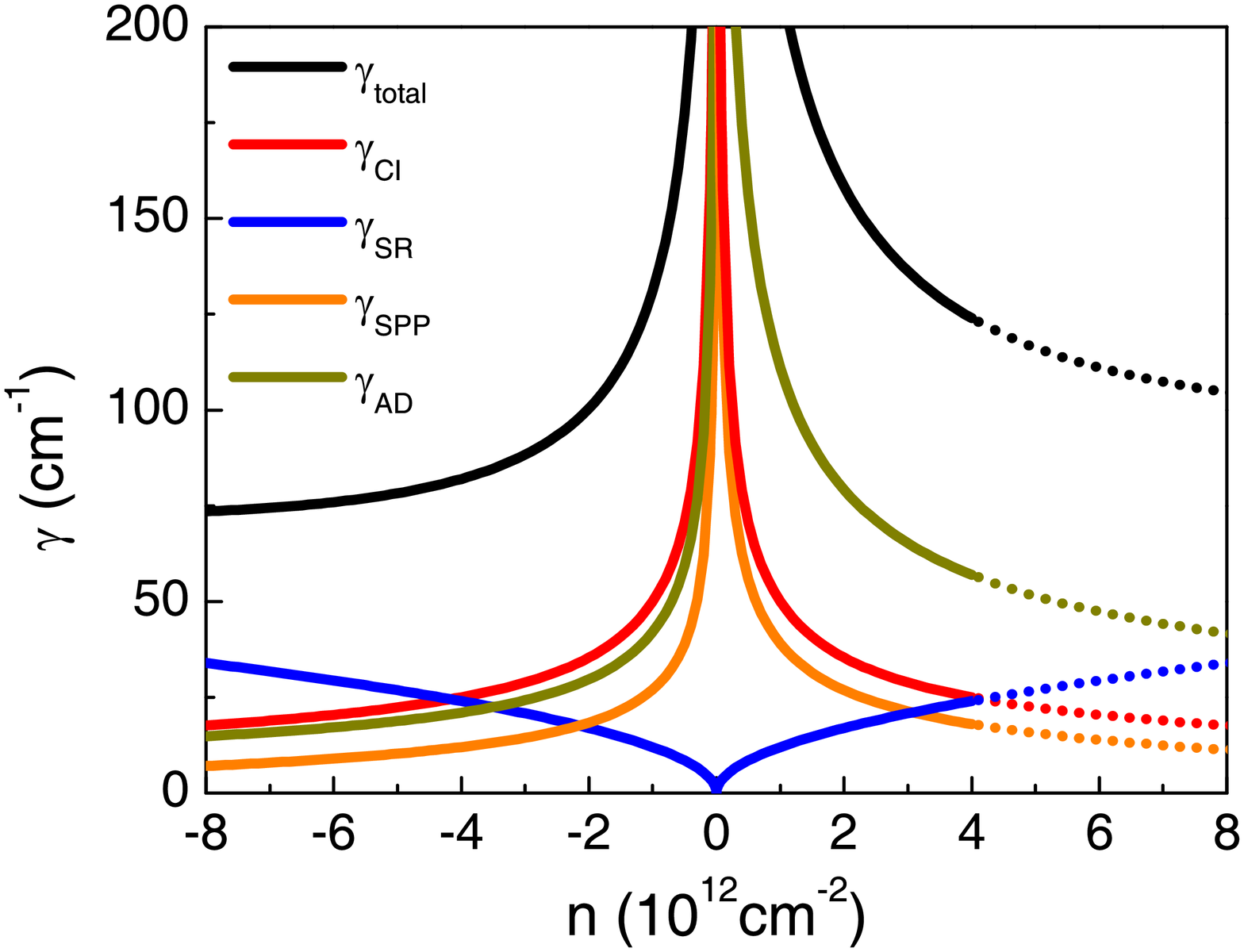}
\caption{
The four scattering sources and their $\gamma(n)$'s identified  and measured in this work: 
$\gamma_{\rm{CI}}$ for charged impurity,
$\gamma_{\rm{SR}}$ for short-range disorder,
$\gamma_{\rm{SPP}}$ for surface polar phonon of SiO$_{2}$, and
$\gamma_{\rm{AD}}$ for gas-adsorbate in air. 
The total scattering $\gamma$ = $\gamma_{\rm{CI}}+\gamma_{\rm{SR}}+\gamma_{\rm{SPP}}+\gamma_{\rm{AD}}$ is shown by the black curve.
}
\label{fig6}
\end{figure*}

\section*{Acknowledgements}
This work was supported by the National Research Foundation of Korea(NRF) grant funded by the Korea government(MSIP)
(NRF-2014R1A2A2A01003448 and NRF-2014R1A2A
2A01006776 (E.H.H.)).

\end{document}